\documentclass[twocolumn,showpacs,preprintnumber,amsmath,amssymb]{revtex4}
\usepackage{dcolumn}
\usepackage{bm}

\newcommand{\pot}[1]{{\phi (x_{#1})}}
\newcommand{\lf}[1]{{\tilde{h}_{#1}}}
\newcommand{\ilf}[1][-10]{{\hat{h}_{#1}}}
\newcommand{\sz}{x_{0}}
\newcommand{\sm}{x_{-1}}
\newcommand{\dsum}[2][ ]{{\displaystyle\sum_{#2}^{#1}}}
\newcommand{\pat}[2][ ]{{\xi^{#2}_{#1}}}

\newcommand{\avn}[1]{{\langle #1 \rangle_{N}}}
\newcommand{\avnp}[1]{{\langle #1 \rangle_{N+2}}}
\newcommand{\rmd}{\mathrm{d}}
\newcommand{\Or}{O}

\begin{document}
\preprint{PRE}
\title{Thouless-Anderson-Palmer equation and self-consistent 
signal-to-noise analysis\\ for the Hopfield model with three-body interaction}
\author{Akihisa Ichiki}
\email{aichiki@mikan.ap.titech.ac.jp}
\author{Masatoshi Shiino}
\affiliation{Department of Applied Physics, Faculty of Science,
Tokyo Institute of Technology, 2-12-1 Ohokayama Meguro-ku Tokyo, Japan}
\date{\today}
\begin{abstract}
The self-consistent signal-to-noise analysis (SCSNA) is an alternative to 
the replica method for deriving the set of order parameter equations for 
associative memory neural network models and is closely related with the 
Thouless-Anderson-Palmer equation (TAP) approach. In the recent paper by 
Shiino and Yamana the Onsager reaction term of the TAP equation has been found 
to be obtained from the SCSNA for Hopfield neural networks with 2-body 
interaction. We study the TAP equation for an associative memory stochastic 
analog neural network with 3-body interaction to investigate the structure of 
the Onsager reaction term, in connection with the term proportional to the 
output characteristic to the SCSNA. We report the SCSNA framework for analog 
networks with 3-body interactions as well as a novel recipe based on the 
cavity concept that involves two cavities and the hybrid use of the SCSNA 
to obtain the TAP equation. 
\end{abstract}
\maketitle

The replica method \citep{SK} for random spin systems has 
been successfully employed in neural network models of associative memory 
\citep{MPV,AGS}. However the replica calculations require the concept 
of free energy. On the other hand an alternative approach to obtain the 
order parameter equations of the self-consistent signal-to-noise analysis 
(SCSNA) \citep{Shiino1,Shiino2} for deterministic analog neural 
networks is free from the energy concept and thus applicable to networks 
with asymmetric connections. The SCSNA was shown to be closely related to 
the TAP equation approach \citep{MPV,TAP,Morita,Shamui,Mezard} 
through the cavity concept in the case where systems have an energy 
function \citep{Shiino3}. 

An advantage of dealing with the TAP equation of neural networks is that 
equilibrium behaviors of a stochastic neural network can be studied by 
investigating the corresponding TAP equation that is viewed as a 
deterministic analog network \citep{Shiino2}. The set of order parameter 
equations of the original stochastic network is obtained by applying the 
SCSNA to the TAP equation. 

TAP equations, which have recently been attracting much attention from the 
viewpoint of applying statistical mechanics to information theoretic 
engineering problems \citep{Jordan,Opper}, are usually obtained by 
means of the Plefka method \citep{Plefka}. A more systematic method of 
deriving TAP equations incorporates the cavity concept \citep{MPV,Shamui} to 
elucidate the structure as well as the meaning of TAP equations. Noting that 
the SCSNA and the TAP equation approach share a common idea of the cavity 
concept, Shiino and Yamana have found that the Onsager reaction term 
characteristic to a TAP equation can be recovered by the SCSNA for stochastic 
analog networks with 2-body interactions \citep{Shiino3}. The aim of the 
present paper is studying the SCSNA framework and the TAP equation of 
stochastic analog networks with multibody interaction to elucidate the 
structure of the Onsager reaction term in connection with the term 
proportional to the output characteristic to the SCSNA. 

The Ising spin Hopfield models with $p$-body connections \citep{Abbott} which 
are analogous to $p$-spin random spin glass models \citep{Rieger} were studied 
to explore the statistical behavior of their retrieval properties. It is 
wellknown that the storage capacity of the network is proportional to 
$N^{p-1}$ \citep{Abbott} where $N$ represents the number of neurons. In the 
present paper we report a novel method based on the cavity approach 
\citep{MPV} to derive the TAP equations for the $p=3$ Hopfield model with 
simultaneous use of the SCSNA for this model. 
Since we deal with analog neurons, or soft spins in the present paper, we 
can choose transfer functions of arbitrary shapes. The TAP equations for 
analog networks with multibody interactions have not been reported. 

We deal with a stochastic analog neural network of the form 
\begin{eqnarray}
\dfrac{\rmd x_{i}}{\rmd t} = -\dfrac{\rmd\pot{i}}{\rmd x_{i}} + 
\dsum{j<k(\neq i)}J_{ijk}x_{j}x_{k} + \eta_{i}(t)\;,\label{dynamics}
\end{eqnarray}
where $x_{i}$ represents a state of an analog neuron or a soft spin at site 
$i$, $\pot{i}$ the potential, $\eta_{i}$ the Langevine white noise obeying 
$\left\langle\eta_{i}(t)\eta_{j}\left( t^{\prime}\right)\right\rangle = 
\frac{2}{\beta}\delta\left( t-t^{\prime}\right)\delta_{ij}$ for 
$i=1\cdots\tilde{N}$, $\beta$ the intensity 
of externally driven Langevin noise and the synaptic coupling $J_{ijk}$ is 
assumed to be given by the Hebb learning rule extended to the $p=3$ Hopfield 
model \citep{Abbott}: 
\begin{equation}
J_{ijk} = \dfrac{1}{\tilde{N}^{2}}\dsum[\tilde{p}]{\mu =1}\pat[i]{\mu}
\pat[j]{\mu}\pat[k]{\mu}
\end{equation}
for $i\neq j\neq k$ and otherwise $J_{ijk} = 0$ with $\pat[i]{\mu}=\pm 1$ 
representing $\tilde{p} (=\alpha\tilde{N}^{2})$ random memory patterns. 

For the $p=3$ Hopfield model, the local field at site $i$ is defined as 
$h_{i}= \sum_{j<k (\neq i)}J_{ijk}x_{j}x_{k}$ and we have to calculate the 
second moments of soft spins $\langle x_{j}x_{k}\rangle$ to obtain the TAP 
equation which is given by expressing the thermal averages of the 
local fields in terms of those of soft spins. For this reason the standard 
cavity method applied to networks with $p=2$, where it suffices to take only 
one cavity into account, is ineffective in the present case. To study the TAP 
equation for the system (\ref{dynamics}) we have to introduce two cavities. 
Hereafter we will refer to the sites $i=0,-1$ as the cavity sites.

The Fokker-Planck equation for the probability density of soft spins 
corresponding to the set of Langevin equations (\ref{dynamics}) with 
$\tilde{N}=N+2$ has the equilibrium probability density 
$P_{\mathrm{eq}}\left( x_{-1},x_{0},x_{1},\cdots ,x_{N}\right) = 
\frac{1}{Z_{N+2}}\exp\left( -\beta H^{(N+2)}\right)$, 
where $Z_{N+2}$ is the partition function of the system including two 
cavities and $H^{(N+2)}$ is its Hamiltonian. 
Noting that the Hamiltonian of $N$-body system $\{ x_{1},\cdots ,x_{N}\}$ is 
given as 
\begin{equation}
H^{(N)} = \dsum[N]{i=1}\pot{i} - \dsum[N]{i<j<k=1} J_{ijk}x_{i}x_{j}x_{k}\;,
\end{equation}
the total Hamiltonian $H^{(N+2)}$ of the system which includes the effects 
of cavity neurons is expressed as 
\begin{eqnarray}
H^{(N+2)} = H^{(N)} + \pot{0} + \pot{-1}\nonumber\\
 - \lf{0}\sz - \lf{-1}\sm - \ilf \sz\sm\;,
\end{eqnarray}
where $\lf{0}$, $\lf{-1}$ and $\ilf$ are respectively defined as
\begin{eqnarray}
\lf{0} = \dfrac{1}{2}\dfrac{1}{(N+2)^{2}}\dsum{\mu} 
\dsum[N]{\substack{j,k = 1\\ j\neq k}}
\pat[0]{\mu}\pat[j]{\mu}\pat[k]{\mu}x_{j}x_{k}\;,\\
\lf{-1} = \dfrac{1}{2}\dfrac{1}{(N+2)^{2}}\dsum{\mu} 
\dsum[N]{\substack{j,k=1\\ j\neq k}}
\pat[-1]{\mu}\pat[j]{\mu}\pat[k]{\mu}x_{j}x_{k}\;,\\
\ilf = \dfrac{1}{(N+2)^{2}}\dsum{\mu} \dsum[N]{k=1} \pat[-1]{\mu}
\pat[0]{\mu}\pat[k]{\mu}x_{k}.
\end{eqnarray}
Notice that $\lf{0}$ and $\lf{-1}$ are the local fields induced by the $N$ 
neurons at cavity sites $i=0,-1$ respectively and $\ilf$ is the local field 
of another type affecting the pair of cavities. 
Then using the marginal probability density and noting that the marginal 
probability density of the local fields of $N$-body system 
is expected to be a three-dimensional Gaussian distribution since 
$\lf{0}$, $\lf{-1}$ and $\ilf$ are the sums of independent random variables, 
one can evaluate the $(N+2)$-body average of the cavity soft spin $x_{0}$ as 
\begin{eqnarray}
\avnp{x_{0}} &=& F\left(\avn{\lf{0}}\right) 
+ \left(\beta\avn{\ilf} + \beta^{2}\sigma_{-10}^{2}\right)\nonumber\\
& &\times\left[ G\left(\avn{\lf{0}}\right)-F^{2}
\left(\avn{\lf{0}}\right)\right] F\left(\avn{\lf{-1}}\right)\nonumber\\ 
& &+ \Or\left( 1/N\right)\;,\label{18}
\end{eqnarray}
where $F$ is the transfer function which is defined as 
\begin{eqnarray}
F(y) &\equiv & \int\rmd x x\exp\Bigl[ -\beta\phi (x)+\beta yx + 
\frac{\beta^{2}\sigma^{2}x^{2}}{2}\Bigr] \nonumber\\
& &\hspace{-15mm}\times\left\{\int\rmd x \exp\Bigl[ -\beta\phi (x)+\beta yx + 
\frac{\beta^{2}\sigma^{2}x^{2}}{2}\Bigr] \right\}^{-1}\label{F}
\end{eqnarray}
and 
\begin{eqnarray}
G(y) &\equiv & \int\rmd x x^{2}\exp\Bigl[ -\beta\phi (x)+\beta yx + 
\frac{\beta^{2}\sigma^{2}x^{2}}{2} \Bigr] \nonumber\\
& &\hspace{-15mm}\times\left\{\int\rmd x \exp\Bigl[ -\beta\phi (x)+\beta yx + 
\frac{\beta^{2}\sigma^{2}x^{2}}{2}\Bigr]\right\}^{-1} ,\label{G}
\end{eqnarray}
$\sigma^{2}$ the variance of the local field $\tilde{h}_{i}$, 
$\sigma_{-10}^{2}$ the covariance of $\tilde{h}_{0}$ and $\tilde{h}_{-1}$ and 
$\langle\cdot\rangle_{N}$ the $N$-body average.
Since $\avn{\ilf} = \Or (1/\sqrt{N})$, $\sigma_{-10}^{2} = \Or (1/\sqrt{N})$, 
\begin{eqnarray}
\avnp{\lf{0}} = \avn{\lf{0}} + \beta\sigma^{2}\avnp{x_{0}} + 
\Or\left( 1/\sqrt{N}\right) ,\label{1}
\end{eqnarray}
and the true local field $h_{0}$ is given by $h_{0} = \lf{0} + \ilf x_{-1}$, 
we have 
\begin{equation}
\avnp{x_{0}} = F\left(\avnp{h_{0}} - \beta\sigma^{2}\avnp{x_{0}}\right)
.\label{pre}
\end{equation}
This equation coincides with the one obtained using the standard cavity method 
for the Hopfield model where one cavity is taken. 

However, it does not suffice to obtain the TAP equation in the present case, 
because $\avnp{h_{0}}$ is simply given in terms of the second 
moments of soft spins $\avnp{x_{j}x_{k}}$. In order to 
find the expression of the local field $\avnp{h_{i}}$ in terms of the 
thermal averages of soft spins, it will be necessary to employ 2-cavity 
method and we obtain 
\begin{eqnarray}
& &\avnp{x_{0}x_{-1}} = \avnp{x_{0}}\avnp{x_{-1}}\nonumber\\
& &\quad + \left(\beta\avnp{\ilf}+ \beta^{2}\sigma_{-10}^{2}\right)\nonumber\\
& &\times\left( \avnp{x^{2}_{0}} - \avnp{x_{0}}^{2}\right)
\left( \avnp{x_{-1}^{2}} - \avnp{x_{-1}}^{2}\right)\nonumber\\
& &\qquad\qquad\qquad\qquad\qquad\quad + \Or\left( 1/N\right) .
\end{eqnarray}
Thus the thermal average of the local field $\avnp{h_{i}}$ is given as 
\begin{eqnarray}
\avnp{h_{i}} = \dsum{j<k(\neq i)}J_{ijk}\avnp{x_{j}}\avnp{x_{k}}\nonumber\\
 + \dfrac{\alpha}{2\beta}\avnp{x_{i}}U^{2} + \Or\left( 1/\sqrt{N}\right)\;,
\label{field}
\end{eqnarray}
where $U$ is the average of susceptibilities 
\begin{eqnarray}
U = \dfrac{\beta}{N+2}\dsum{j}\left(\avnp{x^{2}_{j}} - 
\avnp{x_{j}}^{2}\right) .\label{U}
\end{eqnarray}
Notice that $\avn{\hat{h}_{ij}}$ affects the $(N+2)$-body average of the local 
field $\avnp{h_{i}}$ via the summation with respect to site index, on the 
other hand $\sigma_{-10}^{2}$ does not affect the local field. 
Eq.(\ref{field}) together with Eq.(\ref{pre}) yields the pre-TAP 
equations \citep{Shiino3} for $p=3$ Hopfield model; 
\begin{eqnarray}
\avn{x_{i}} = F\left(\dsum{j<k(\neq i)}J_{ijk}\avn{x_{j}}\avn{x_{k}} - 
\Gamma_{\mathrm{TAP}}\avn{x_{i}}\right)\nonumber\\
\hspace{-15mm}(i=1,\cdots ,N)\label{pTAP}
\end{eqnarray}
with 
\begin{eqnarray}
\Gamma_{\mathrm{TAP}} = \beta\sigma^{2} - \dfrac{\alpha U^{2}}{2\beta},
\label{32}
\end{eqnarray}
where $\avn{x_{i}}$ is used instead of $\avnp{x_{i}}$. $-\Gamma_{\mathrm{TAP}}
\avn{x_{i}}$ is the Onsager reaction term of the TAP equation. However, 
$\Gamma_{\mathrm{TAP}}$, especially, $\sigma^{2}$ in Eq.(\ref{F}) has 
still to be determined to obtain the TAP equation. 

Eq.(\ref{pTAP}) defines a deterministic analog network of associative 
memory with the transfer function $F$ corresponding to the original 
stochastic analog neural network (\ref{dynamics}). To obtain $\sigma^{2}$ 
together with the order parameter equations we then apply the SCSNA to 
Eq.(\ref{pTAP}). The SCSNA is a self-consistent method for properly 
renormalizing the so-called noise part due to interference of noncondensed 
patterns in the local field of a neuron 
$h_{i}^{\mathrm{det}}\equiv\sum_{j<k(\neq i)}J_{ijk}\avn{x_{j}}\avn{x_{k}}$ 
in Eq.(\ref{pTAP}), which can be rewritten as 
\begin{eqnarray}
h_{i}^{\mathrm{det}} =\dfrac{1}{2}\displaystyle\sum_{\mu =1}^{\alpha N^{2}}
\xi_{i}^{\mu}\left[ (m^{\mu})^{2} - \dfrac{1}{N}\sum_{j}
\dfrac{\avn{x_{j}}^{2}}{N}\right]\nonumber\\
- \dfrac{\avn{x_{i}}}{N}\sum_{\mu =1}^{\alpha N^{2}}m^{\mu} + 
\dfrac{\avn{x_{i}}^{2}}{N^{2}}\sum_{\mu =1}^{\alpha N^{2}}\xi_{i}^{\mu}\;,
\end{eqnarray}
where $m^{\mu}$ is the overlap of $\mu^{\mathrm{th}}$ pattern defined as 
$m^{\mu} \equiv \frac{1}{N}\sum^{N}_{i=1}\pat[i]{\mu}\avn{x_{i}}$ 
for $\mu = 1,\cdots ,\alpha N^{2}$. To extract the pure noise obeying a 
Gaussian distribution, we decompose the local field $h_{i}^{\mathrm{det}}$ 
assuming that the first pattern is condensed and the others are noncondensed, 
i.e., $m^{1}=\Or (1)$ and $m^{\mu}=\Or (1/\sqrt{N})$ ($\mu\ge 2$) 
as \citep{Shiino1,Shiino2}
\begin{eqnarray}
h_{i}^{\mathrm{det}} = \dfrac{1}{2}\pat[i]{1}\left( m^{1}\right)^{2} + 
\dfrac{1}{2}\pat[i]{\mu} \left[ \left( m^{\mu}\right)^{2} - 
\dfrac{\tilde{C}}{N}\right] - \dfrac{\avn{x_{i}}}{N}m^{\mu}\nonumber\\
\hspace{-25mm} + \dfrac{\avn{x_{i}}^{2}}{N^{2}}\xi_{i}^{\mu} 
+ z_{i\mu} + \Gamma_{\mathrm{SCSNA}} \avn{x_{i}}\;,\label{local}
\end{eqnarray}
where
\begin{eqnarray}
\tilde{C} &\equiv& \dfrac{1}{N}\dsum[N]{i=1}\avn{x_{i}}^{2}\;,\\
\Gamma_{\mathrm{SCSNA}} &\equiv& \gamma - \tilde{\gamma}\;,\\
\tilde{\gamma} &\equiv& \dfrac{1}{N}\dsum{\mu \ge 2}m^{\mu}\;,\\
z_{i\mu} + \gamma \avn{x_{i}} &\equiv& \dfrac{1}{2}\dsum{\nu\neq 1,\mu}
\pat[i]{\nu}\left[ \left( m^{\nu}\right)^{2} - \dfrac{\tilde{C}}{N}\right] ,
\end{eqnarray}
and $z_{i\mu}$ is assumed to be a Gaussian random variable with mean zero 
and the variance is to be evaluated self-consistently. 

Substituting the expression of the local field Eq.(\ref{local}) into the 
pre-TAP equation (\ref{pTAP}) and comparing with $\avn{x_{i}} = 
F\left(\langle\tilde{h}_{i}\rangle_{N-2}\right)$, which holds by observing 
Eq.(\ref{18}) for large $N$, it follows that 
$\Gamma_{\mathrm{SCSNA}} = \Gamma_{\mathrm{TAP}}$ \citep{Shiino3}, since 
$\langle\tilde{h_{i}}\rangle_{N-2} - \frac{1}{2}\xi_{i}^{1}(m^{1})^{2}$ 
obeys a Gaussian random variable. Then we have 
\begin{eqnarray}
\avn{x_{i}} = F\left(\dfrac{1}{2}\pat[i]{1}\left( m^{1}\right)^{2} 
+ \dfrac{1}{2}\pat[i]{\mu}\left[ \left( m^{\mu}\right)^{2} - 
\dfrac{\tilde{C}}{N}\right] + z_{i\mu}\right)
\end{eqnarray}
from which it follows \citep{Shiino2} 
\begin{eqnarray}
m^{\mu} &=& \dfrac{1}{N}\dsum[N]{i=1}\pat[i]{\mu}
F\left(\dfrac{1}{2}\pat[i]{1}\left( m^{1}\right)^{2} 
+ z_{i\mu}\right)\nonumber\\
&+& \dfrac{1}{2}\left[ (m^{\mu})^{2} - \dfrac{\tilde{C}}{N}\right]
\dfrac{1}{N}\dsum[N]{i=1}F^{\prime}\left(\dfrac{1}{2}\pat[i]{1}
\left( m^{1}\right)^{2} + z_{i\mu}\right)\nonumber\\
& & + \Or\left( N^{-5/2}\right) \;,\label{ptb}
\end{eqnarray}
where $F^{\prime}$ denotes the derivative of the transfer function $F$. 

{\it It should be noted that in the present case it is necessary to obtain }
$m^{\mu}(=\Or (1/\sqrt{N}))$ {\it up to }$\Or (1/N)$ {\it unlike the case of }
$p=2$ {\it where up to }$\Or (1/\sqrt{N})$ {\it of }$m^{\mu}$ {\it suffices}. 
We solve this equation for $m^{\mu}$ perturbatively by putting 
$m^{\mu} = m^{\mu}_{1/2} + m^{\mu}_{1} + \Or\left( N^{-3/2}\right)$, 
where $m^{\mu}_{k}$ represents the part of $\Or (N^{-k})$. 
Substituting the solution of Eq.(\ref{ptb}) into the definitions of 
$\Gamma_{\mathrm{SCSNA}}$ and $z_{i\mu}$, we find $\tilde{\gamma}=0$ in the 
limit of large $N$. 
Furthermore, $\Gamma_{\mathrm{SCSNA}}$ which determines the form of the 
transfer function $F$ by the relation 
\begin{eqnarray}
\Gamma_{\mathrm{SCSNA}} = \Gamma_{\mathrm{TAP}} = 
\beta\sigma^{2} - \frac{\alpha U^{2}}{2\beta} \label{gammaS}
\end{eqnarray}
is evaluated using the self-averaging property as 
\begin{eqnarray}
\Gamma_{\mathrm{SCSNA}} = \alpha U\left\langle F^{2}\left(\dfrac{1}{2}
\pat{1}\left( m^{1}\right)^{2} + z\right)\right\rangle_{\pat{1},z}\;,
\label{gamma}
\end{eqnarray}
where $U$ is given as 
\begin{eqnarray}
U = \left\langle F^{\prime}\left(\dfrac{1}{2}\pat{1}
\left( m^{1}\right)^{2} + z\right)\right\rangle_{\pat{1}, z}\label{Uself}
\end{eqnarray}
and $\langle\cdot\rangle_{\pat{1}, z}$ represents the average over the 
random pattern $\pat{1} = \pm 1$ and the Gaussian random variable $z$. 
Notice that $U$ given by this equation corresponds to that defined by 
Eq.(\ref{U}), since differentiating Eq.(\ref{F}) with respect to $y$ yields 
the susceptibility and the self-averaging property holds. 
Similarly the variance of the Gaussian noise $z$ can be calculated as 
\begin{eqnarray}
\sigma_{z}^{2} = \dfrac{\alpha}{2}\left\langle F^{2}\left(\dfrac{1}{2}
\pat{1}\left( m^{1}\right)^{2} + z\right) \right\rangle_{\pat{1},z}^{2}.
\label{sigma}
\end{eqnarray}
Furthermore the overlap of the condensed pattern $m^{1}$ is given as 
\begin{eqnarray}
m^{1} = \left\langle\pat{1}F\left(\dfrac{1}{2}\pat{1}
\left( m^{1}\right)^{2} + z\right)\right\rangle_{\pat{1},z}.\label{condm}
\end{eqnarray}
Eqs.(\ref{Uself}), (\ref{gamma}), (\ref{sigma}) and (\ref{condm}) constitute 
the SCSNA framework in terms of $m^{1}$, $U$, $\sigma^{2}$, and 
$\sigma_{z}^{2}$ that yields the order parameter equations for determining 
the storage capacity of the analog network 
(\ref{pTAP}) and hence of the stochastic analog neural 
network (\ref{dynamics}). We note that they coincide with those obtained by 
the replica calculation when the replica symmetry is assumed to hold. 
Furthermore, in the special case where $\exp -\beta\phi (x)/\int\exp -\beta
\phi (x)\mathrm{d}x = \frac{1}{2}\left(\delta (x+1)+\delta (x-1)\right)$, 
the result of the SCSNA recovers that of the Ising spin network studied in 
\cite{Abbott}. Then, 
from Eqs.(\ref{pTAP}) and (\ref{32}) we finally obtain the TAP equation as 
\begin{eqnarray}
\avn{x_{i}} = F\hspace{-1mm}\left(\dsum{j<k(\neq i)}J_{ijk}\avn{x_{j}}
\avn{x_{k}} - \Gamma_{\mathrm{SCSNA}}\avn{x_{i}}\hspace{-1mm}\right)
\label{TAPeq}
\end{eqnarray}
with $\Gamma_{\mathrm{SCSNA}}$ and the form of the transfer function $F$ 
self-consistently determined by the above Eqs.(\ref{Uself}), 
(\ref{gammaS}), (\ref{gamma}), (\ref{sigma}) and (\ref{condm}). Noting 
that $\Gamma_{\mathrm{SCSNA}}$ can be expressed in terms of Edwards-Anderson 
order parameter $q$, the TAP equation (\ref{TAPeq}) can be further rewritten 
as $\avn{x_{i}} = F(\sum_{j<k(\neq i)}J_{ijk}\avn{x_{j}}
\avn{x_{k}} - \alpha\beta q(\hat{q} - q)\avn{x_{i}})$, 
where the Edward-Anderson order parameter $q$ and $\hat{q}$ are defined 
respectively as 
$q = \frac{1}{N}\sum_{i}\avn{x_{i}}^{2}$ and 
$\hat{q} = \frac{1}{N}\sum_{i}\avn{x_{i}^{2}}$ which are easily computed 
using Eqs.(\ref{F}) and (\ref{G}).

In conclusion, taking advantage of the close relationship between the TAP 
equation and the SCSNA approaches, we have studied the SCSNA framework and 
the TAP equation for the stochastic analog network with 3-body interaction 
based on the Hebb learning rule. Characteristic to the TAP equation for the 
case of such interaction is that the Onsager reaction term consists of the 
two terms, as is shown in Eq.(\ref{32}): the one arising from the variance of 
the Gaussian distribution for the local fields that determines the shape of 
the transfer function $F$ and the other one due to the two body correlation 
of soft spins the computation of which requires taking two cavities. 
Without resorting to the Hamiltonian based on overlap evaluation by adding a 
memory pattern to the network, which is usually taken for the Hopfield model 
\cite{MPV,Shiino3}, the coefficient of the Onsager reaction term has been 
given within the framework of the SCSNA, although applying the conventional 
recipe recovers the result. Renormalization of the noise part of the local 
field in the SCSNA scheme corresponds to the overlap evaluation in the 
pattern-adding approach of the cavity method. The framework 
of the SCSNA described in its application to the (pre-)TAP equation of our 
stochastic network implies that the set of order parameter equations obtained 
there still formally makes sense in the case of general deterministic analog 
networks having 3-body interaction and transfer function $F$ of arbitrary 
shape irrespective of whether it is monotonic or not. Details of the analysis 
including the phase diagram showing the behavior of the storage capacity will 
be published elsewhere. 

This work was supported by a 21st Century COE Program 
at TokyoTech "Nanometer-Scale Quantum Physics" by the Ministry of Education, 
Culture, Sports, Science and Technology. 

\end{document}